\documentclass[sigconf]{acmart}

\AtBeginDocument{%
  \providecommand\BibTeX{{%
    \normalfont B\kern-0.5em{\scshape i\kern-0.25em b}\kern-0.8em\TeX}}}

\copyrightyear{2021}
\acmYear{2021}
\setcopyright{acmcopyright}\acmConference[CHI '21]{CHI Conference on Human Factors in Computing Systems}{May 8--13, 2021}{Yokohama, Japan}
\acmBooktitle{CHI Conference on Human Factors in Computing Systems (CHI '21), May 8--13, 2021, Yokohama, Japan}
\acmPrice{15.00}
\acmDOI{10.1145/3411764.3445649}
\acmISBN{978-1-4503-8096-6/21/05}

\begin{document}

\title{Towards an Understanding of Situated AR Visualization for Basketball Free-Throw Training}


\author{Tica Lin, Rishi Singh, Yalong Yang, Carolina Nobre, Johanna Beyer, Maurice A. Smith, \\ Hanspeter Pfister
}
\affiliation{Harvard University, Cambridge, MA, USA}

\email{{ mlin, rishibalsingh, yalongyang, cnobre, jbeyer, mas, pfister } @g.harvard.edu} 

\renewcommand{\shortauthors}{Lin et al.}


\begin{abstract}


We present an observational study to compare co-located and situated real-time visualizations in basketball free-throw training. Our goal is to understand the advantages and concerns of applying immersive visualization to real-world skill-based sports training and to provide insights for designing AR sports training systems. 
We design both a situated 3D visualization on a head-mounted display and a 2D visualization on a co-located display to provide immediate visual feedback on a player's shot performance. Using a within-subject study design with experienced basketball shooters, we characterize user goals, report on qualitative training experiences, and compare the quantitative training results. 
Our results show that real-time visual feedback helps athletes refine subsequent shots. Shooters in our study achieve greater angle consistency with our visual feedback. 
Furthermore, AR visualization promotes an increased focus on body form in athletes.
Finally, we present suggestions for the design of future sports AR studies.

\end{abstract}


\begin{teaserfigure}
  \includegraphics[width=\textwidth]{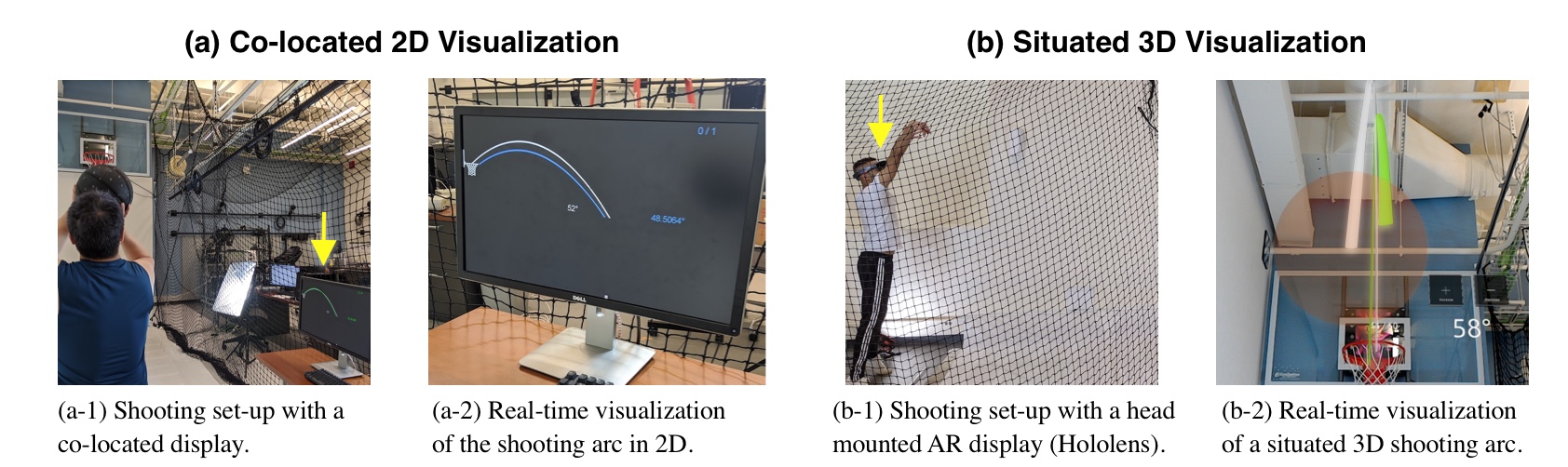}
  \caption{Visualization design for real-time visual feedback during basketball free-throw shot practice. (a) A co-located 2D visualization on a desktop display (yellow arrow). (b) First-person view of a situated 3D visualization on an AR HMD.}
  \Description[Visualization design for real-time visual feedback during basketball free-throw shot practice.]{Visualization design for real-time visual feedback during basketball free-throw shot practice. (a) A co-located 2D visualization on a desktop display (yellow arrow). (b) First-person view of a situated 3D visualization on an AR HMD.}
  \label{fig:shooting_arc}
  \vspace{4mm}
\end{teaserfigure}


\maketitle


\section{Introduction}

Learning and perfecting a new skill takes practice.
For example, playing sports or a musical instrument requires developing and refining specific motor control skills to increase accuracy, consistency, speed, and timing. 
Athletes, as well as musicians, need to learn new movement patterns, integrate those new patterns with previously learned movements, and learn proper sequencing of those patterns.
Motor skill development is a crucial part of sports practice, next to other factors such as team strategy, endurance, or scene understanding.
The quality of practice greatly influences the pace of improvement. 
Therefore, efficient practice and training typically require personalized coaching and evaluation.
Complex motor skill learning, in particular, requires \emph{frequent and immediate feedback}~\cite{frequent_feedback_motor_skill}.

In recent years, sports analytics~\cite{alamar_sports_2013} has gained popularity in sports management, coaching, and training, by embedding data-driven decision making into the training process. 
Motion capture systems can now collect player-specific data, which can be used in visual analytic systems~\cite{Perin2018-jh} for detailed performance analysis~\cite{ye2020shuttlespace}. 
However, most of these systems do not provide \emph{immediate} feedback during practice, which is crucial for rapid skill improvement~\cite{frequent_feedback_motor_skill, KR_frequency}. 
To augment sports skill training with immediate feedback, co-located and situated displays are both pragmatic approaches. HomeCourt~\cite{homecourt} and Noah~\cite{noah} are two commercial applications that track a basketball player's movement to give shot feedback. However, they do not give immediate \emph{visual} feedback to athletes but focus on auditory feedback or rely on the presence of coaches. Additionally, their effectiveness has not been formally evaluated.

With the emergence of head-mounted displays (HMD), augmented reality (AR) can now provide \emph{immediate and situated visual feedback} in real-world training settings, potentially further shortening the feedback loop and speeding up motor skill learning.
HMDs support the embedding of virtual data into the real world, have relatively low hardware costs, and support hands-free user input, all crucial for sports training.
While there are some user studies evaluating training with situated AR visualization in some sports, like dancing~\cite{chan2010virtual,kyan2015approach}, climbing~\cite{kajastila2016augmented}, hand ball~\cite{jensen_keepin_2015}, and table tennis~\cite{altimira_digitally_2016}, to the best of our knowledge 
and based on recent surveys of AR applications~\cite{fengzhou_2008_Trends,dey_systematic_2018,kim_revisiting_2018,soltani_2020_AugmentedRealitya}, no previous work has evaluated situated visualizations for motor skill training in basketball with real athletes. 

In this paper, we examine how athletes use situated AR visualization in realistic sports training and what aspects of augmented information are useful.
We aim to characterize AR in sports training to answer: \emph{``What are the unique aspects of using AR for sports training?''}, \emph{``What are the advantages and concerns of applying AR to actual sports training?''}, and \emph{``What are the most important aspects to take into account when designing sports AR studies?''}.

This work presents the first study to compare co-located versus situated real-time visualizations in basketball free-throw training.
We collaborated with two National Collegiate Athletic Association (NCAA) basketball teams to design a novel situated visualization system for basketball free-throw shot training. We discovered that players seek real-time feedback to achieve a \emph{better and more consistent shot arc} in their shooting practice.
Therefore, we implemented a co-located 2D and a situated AR visualization with immediate visual feedback on an athlete's shooting arc.
For the 2D condition, we designed a 2D visualization on a standard monitor display that is positioned next to the basketball player during
free-throw practice. For the AR condition, we designed a situated 3D visualization using an HMD that is worn during practice (Fig.~\ref{fig:shooting_arc}).

We present the results of a quantitative and qualitative user study with nine experienced basketball players to characterize how the two visualization conditions affect performance and user goals. Based on the results of our experiments, we can offer evidence in direct support of several guidelines for immersive sports visualizations, including that: (1) participants appreciated real-time visual feedback when gauging their performance and actively used it to refine their subsequent shots, (2) each visualization modality naturally supports different user goals, (3) performance in shot angle consistency improved throughout the study in both, the 2D and AR conditions, (4) the AR condition was initially less familiar to users and required more fine-grained adjustment to each individual,
and (5) the extra spatial information provided by the AR condition provided a more realistic feel and resulted in a more holistic sense of participant performance and body position.


\section{Related Work}
AR technology can render interactive 2D and 3D graphics in any space or surface around the user, which provides possibilities to improve the existing workflows in many fields. 
Two survey papers reviewed AR technologies and applications of the last two decades~\cite{fengzhou_2008_Trends,kim_revisiting_2018}. 
Particularly, Kim et al.~\cite{kim_revisiting_2018} identified maintenance, simulation, and training in industrial, military, and medical fields as the most popular AR application topics. 
The overarching assumption behind these applications is that real-time situated visual guidance or feedback can potentially augment the user's capabilities and subsequently improve the user's skills~\cite{marriott_situated_2018}.
Sports share many characteristics with these popular applications. However, using AR in sports has been less explored~\cite{lin2020sportsxr}. In this paper, we focus on basketball free-throw training. 
There is limited research on this specific topic, so we review papers on providing immediate visual support in general sports training with an emphasis on using AR in basketball free-throw training, when possible. 

Broadly speaking, there are three main approaches for providing immediate visual support in sports training: situated AR, co-located displays, and VR simulations. VR simulations for sports have been extensively studied~\cite{neumann_2018_Systematic,akbas_2019_ApplicationVirtuala,miles_2012_Review,faure_2020_VirtualRealitya}. Although some research demonstrates preliminary positive results of using VR for sports training, many challenges remain for achieving ideal fidelity in VR simulations. 
Specifically, Covaci et al.~\cite{covaci_2015_Visuala} investigated basketball free-throw training in a projector-based VR environment and found that participants underestimated distance in VR.
In our study, we focus on \emph{situated AR} and \emph{co-located display} approaches, where players can see accurate visual information in their physical environment during shot practice. 

\noindent\textbf{Situated AR approaches.}
Basketball and rock climbing are the sports that have received the most attention in the AR literature so far~\cite{soltani_2020_AugmentedRealitya}.
For basketball, approaches include mobile AR games~\cite{hebbel-seeger_2015_Physical,santoso_2018_Markerless}, audio and visual feedback for pickup games~\cite{ryan_2017_2KReality, chatham2013adding}, wearable displays on a player's uniform to display game-related statistics~\cite{page_2007_Evaluating}, and AR games with a virtual ball~\cite{baudisch_imaginary_2013}.
While these systems are relevant to basketball, none of them focuses on motor skill training.

Kahrs et al.~\cite{kahrs_supporting_2006} developed an early AR prototype for providing visual guidance in basketball shooting.
Their system visualizes a 3D trajectory to the center of the rim. However, while it provides visual guidance before a shot, the system does not track and visualize the player's actual shots. 
The player cannot compare the difference between the guided trajectories and their real shots, thus making it difficult to identify what they need to change to improve. Kahrs et al. also did not formally evaluate their system.
Furthermore, the hardware of AR HMDs has improved dramatically since that early prototype, now offering higher resolution, more stable tracking, and a larger Field of View (FoV).
Compared to their system, our system enables real-time basketball tracking using pre-installed cameras. With real-time basketball tracking, we can render the trajectory of the real shot. Thus, the user can visually compare the difference between the guided trajectory and the real shot.
We also systematically evaluated the effectiveness of our system with a state-of-the-art AR HMD (Microsoft HoloLens).

\noindent\textbf{Co-located display approaches.}
Placing a display next or onto a sports field is an easy way to provide immediate visual feedback to athletes. 
A few user studies have demonstrated positive results in different sports.
Crowell et al.~\cite{crowell_reducing_2010} found that plotting real-time tibial acceleration data on a 2D screen placed in front of a treadmill can reduce peak tibial acceleration associated with stress fractures. 
Kaplan et al.~\cite{kaplan_-situ_2016} display pedal pressure overlaid on videos for cycling, and their participants were generally positive about their system. 
Chan et al.\cite{chan_virtual_2011} and Anderson et al.~\cite{anderson_youmove_2013} used a similar idea of presenting the skeleton of a tracked user in front of the user, for dancing and general motion training, respectively. 

Overall, co-located display approaches are well-studied and are likely to be beneficial for specific sports. 
Compared to situated AR approaches, co-located displays may introduce extra context switching costs, i.e., players need extra cognitive efforts to map the visual feedback from the 2D display to the physical space.
We developed a co-located display system with the same basketball tracking setup as our situated AR system. We placed a 2D monitor close to the player to show visual guidance and feedback and to allow players to access real-time visual information on the screen during training. 
We compared this co-located display setup with our situated AR system in a controlled user study.

\noindent\textbf{Applications for basketball free-throw training.}
The Noah Shooting System~\cite{noah} and HomeCourt~\cite{homecourt} are the two most well-known applications for basketball free-throw training, but neither has been quantitatively evaluated.
The Noah Shooting System uses pre-installed cameras and sensors, and HomeCourt uses the camera on a mobile device to track the ball. 
Both systems provide real-time audio feedback about the previous shot, most commonly the captured shooting angle.
The Noah Shooting System does not have any displays on the court, so the player cannot access any visual feedback during training. 
In Homecourt, the mobile device needs to be placed at a distance to capture the full shooting trajectory with its built-in camera.
Therefore, players cannot see the mobile display showing shot data during training. 
In practice, both systems provide either real-time audio feedback or need a second person (e.g., a coach) to access the visual information on a screen.

However, getting audio feedback on a player's numerical shot angle does not necessarily give the player intuitive clues on how to best improve their shot.  
One of our main hypotheses in this work is that real-time visual guidance and feedback may be able to convey more nuanced details than audio feedback alone and may give players more intuitive clues for improvement.

Our situated AR and co-located display systems can provide real-time visual guidance and feedback to the player during training.
In this paper, we investigate the potential benefits of such real-time visual guidance and feedback for basketball free-throw training. 
Whether those potential benefits are transferable (i.e., improvements persist after removing the visual guidance and feedback) is beyond the scope of this paper, and studying transferability is discussed as a future research direction in Sec~\ref{sec:discussion}.


\section{Basketball Free-throw Training}
\label{sec:background}

\subsection{Background}
Unlike regular shots in a basketball game, the free-throw shot is a pure test of skill unfettered by defending players and changing conditions. 
Free-throw shots are awarded when a player is fouled during a shot attempt and allow the player to shoot at the hoop from a distance of 15 feet (i.e., the free-throw line) while all other players stand aside.
Despite its sizable contribution to team offense, scoring, winning games~\cite{free_throw_importance_playoffs_2003, free_throw_importance_game_stage_1994, performance_indicators_winning_2009}, and essentially being "free" points, the free-throw shot in basketball is a difficult skill to improve~\cite{nytimes_freethrow, guardian_freethrow}.
For the last 50 years, the league-average free-throw shot percentages in the NBA and the NCAA have remained nearly static at a mediocre 75\% ~\cite{basketball-reference-nba} and 70\%~\cite{sports-reference-ncaa}.  
Due to its controlled nature and the lack of improvement over several decades, the free-throw shot is a valuable testing ground for strategies to improve motor skill learning.

A likely contributor to the lack of precise tools and methods for improving a player's free-throw shooting is an insufficient understanding of what factors a skilled shooter optimizes to become more consistent and accurate. 
Studies found that the movement of a skilled shooter is less variable from shot-to-shot compared to those less skilled~\cite{Arm_joint_kinematics_2015, 1981_freethrow_kinematics, coordination_variability_miss_vs_swish, nature_neuro_basketball}.
Tran and Silverberg focused on how the ball's release parameters affect the probability of success~\cite{Silverberg_freethrowsimulation}. However, their approach does not account for how humans' natural movement inconsistency may be affected when shooting in a particular way. 
More recently, Nakano et al.~\cite{Nakano_min_release_speed_2020} found that moderately skilled shooters grossly minimize the ball release speed during shooting to reduce motor noise. 
However, they do not take into account individual differences in the shooter's release positions and heights.
Generally, the scientific understanding of optimal shooting is not complete enough to provide evidence-based recommendations on shooting form outside of some small basic factors.

Assuming there is an optimal shooting motion for a given player, improvement in free-throw shooting would require that players achieve the optimal shooting motion on average, but also that they become \emph{more consistent around that average motion}. 
Variability in shooting form is a shot-by-shot phenomenon. 
Therefore, it would be extremely helpful for players to get feedback on each shot's deviation in shooting form.
There is evidence that even reducing the frequency of feedback from every trial to every other trial can significantly slow down learning and reduce retention~\cite{frequent_feedback_motor_skill, KR_frequency, fujii2016more}.

\subsection{Goal and Task Analysis}
\label{sec:goal_task_analysis}

Based on the above literature review on basketball free-throw shooting, there currently is an incomplete scientific understanding of what factors constitute an optimal shooting form. However, it has been shown that real-time feedback on one's shooting form is beneficial for motor learning~\cite{silva2017improved}.
We conducted semi-structured user interviews with several coaches and players on Harvard Women's and Men's Basketball teams to identify the important factors in their free-throw shot training. When asked \textit{``What factors do you look at to evaluate your shooting performance
during practice?''}, coaches focus on watching players' \emph{``shooting form''} while players focus on their \emph{``shot arc.''} In practice, most players practice free-throw shooting alone without coaching feedback. 

The skilled shooters we interviewed, on average, make 80\% of their free-throws and all aim for shooting at a \emph{``perfect arc''} or \emph{``better arc''}. However, only one of them knew the \emph{precise angle} of their target arc. She practices shooting free-throws at a specific entry angle (44-46 degrees) with the Noah training system~\cite{noah}. 
She considers the real-time audio feedback of angle number and after-the-fact shot visual analysis from the Noah system useful to train her shot arc. 
The other players reported to only subjectively evaluate their shot performance by visually following the shot arc to the rim (e.g., identifying that a shot was flat and short). Then they make small adjustments to their shooting forms to change ball direction, spin, release height, and speed according to \emph{``their feeling.''}
All players were highly interested in accessing immediate visual feedback of their shot arc during shot training.

We extracted some main recurring themes and goals for shot training from our user interviews: \textbf{G1 - improving free-throw performance by using quantitative feedback immediately during practice}, \textbf{G2 - increasing shot consistency}, and \textbf{G3 - improving one's shot arc}.
With only auditory or no real-time feedback at all, there is currently a lack of precise goal specification and actionable outcome evaluation.
For example, there is no intuitive way for a player to specify what an ideal arc should look like and subsequently evaluating how much higher the current shot was compared to the previous shot.
Therefore, it is difficult for a player to make an accurate adjustment. 
Furthermore, immediate feedback from coaches is not available most of the time. 
Thus, the goal of our system is to offer intuitive visual feedback for players in real-time without requiring the presence of coaching staff.

Based on our interviews, we have identified the following tasks that our system needs to support in free-shot training:

\noindent
\textbf{T1: Analysis of the target shooting arc before a shot.} 
Currently, players can only visualize the shooting arc they aim for (i.e., target arc) in their minds before they take a shot. 
However, seeing the actual target arc spatially situated in the player's environment immediately before taking a shot would give athletes highly accurate visual guidance.

\noindent
\emph{Challenge - Spatial visualization of shot arcs:} 
Any shooting arc in our system needs to be shown as a visual arc, situated in the player's actual physical space to give depth cues and support three-dimensional reasoning. 
Furthermore, the target arc visualization needs to be intuitive and unobtrusive to not distract the player during their shot.

\noindent
\textbf{T2: Analysis of one's shooting arc during and after each individual throw during practice.}
The most crucial task for basketball players during shot practice is to get detailed spatial information about their shots, directly during their training session. 

\noindent
\emph{Challenge - Real-time visualization and immediate visual feedback directly on the court:} 
Visual feedback should be available during practice and after every shot. 
Specifically, players should not have to pause their practice to analyze their last shot, but have direct hands-free access to visual feedback. Furthermore, the time between the end of the shooter's motion (ball release) and when the visual feedback is available should be minimized. Delays in sensory feedback have been shown to impact motor learning ability~\cite{Brudner2016-yw, Honda2012-gq}.

\noindent
\textbf{T3: Comparing one's actual shooting arc to the target shooting arc.} 
To extract actionable insights from shot feedback, players must be able to compare their latest shot to their set target arc. This allows players to make informed decisions about how to adjust their next shot.

\noindent
\emph{Challenge - Visual comparison with minimal user input:} 
Currently, there is no way for players to get quantitative feedback on how their shot compares to their set target shot. 
Comparing both shot arcs should be immediate, easy to understand, and not require any user input. 

\noindent
\textbf{T4: Adjusting one's personal target arc to get consistently closer to an \emph{ideal} arc.}
The ideal shooting arc varies from player to player, depending on height, physical fitness, and personal preferences. It is crucial for players to be able to set their personal target arc, and to be able to adjust it over weeks or months as they improve their form.

\noindent
\emph{Challenge - Adjustable target shot angle:} 
User input for situated and co-located visualizations, especially in sports settings, should be minimal and not require complicated input devices.

\begin{figure}[b!]
    \centering
    \vspace{-4mm}
    \includegraphics[width=\linewidth]{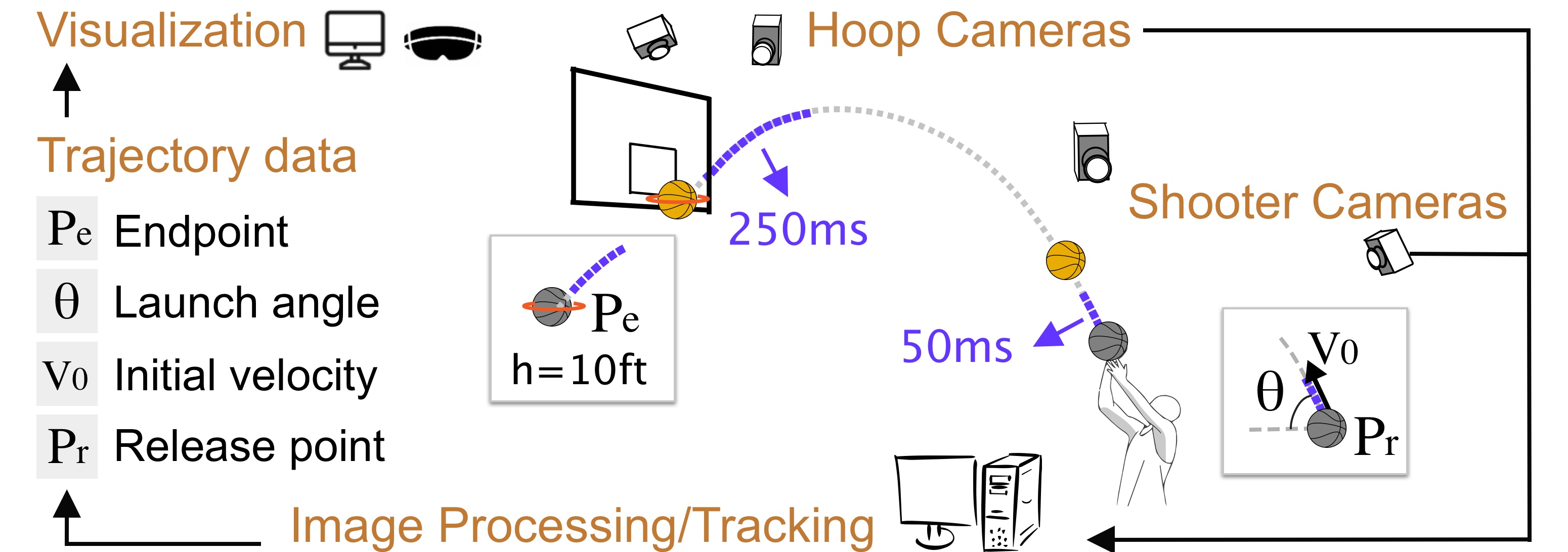}
    \caption{Our 3D motion capture system consists of two shooter and two hoop cameras, which track the first 50 ms upon ball release and the last 250 ms toward the end of the shot, respectively, to calculate launch angle and shot arc.} 
    \Description[The diagram of our 3D motion capture system.]{Our 3D motion capture system consists of two shooter and two hoop cameras, which track the first 50 ms upon ball release and the last 250 ms toward the end of the shot, respectively, to calculate launch angle and shot arc.}
    \label{fig:system}
\end{figure}

\begin{figure*}[tb!]
    \centering
    \includegraphics[width=\linewidth]{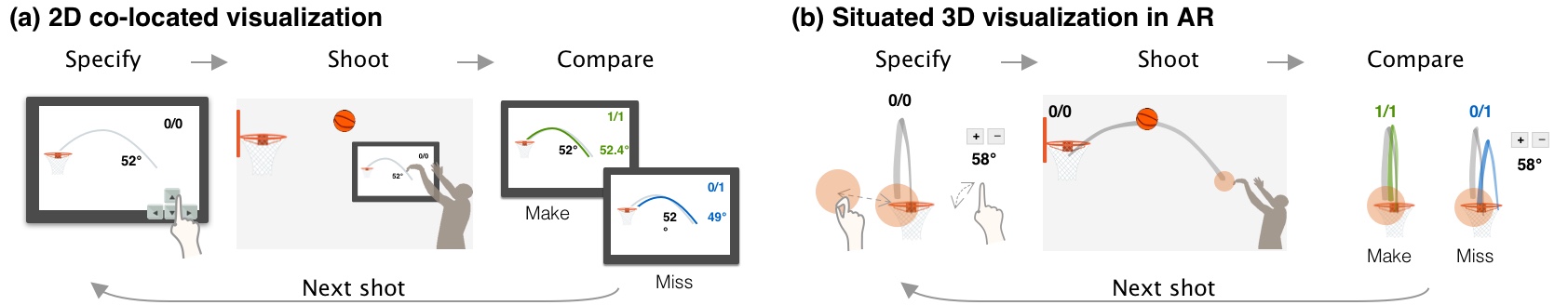}
    \caption{(a) and (b) show the visualization design and interaction loop of the 2D and AR visualizations. Users interactively specify their target shot arc, shoot the ball, and compare their shot arc to the target arc.} 
    \Description[The visualization design and interaction loop of the 2D and AR visualizations.]{(a) and (b) show the visualization design and interaction loop of the 2D and AR visualizations. Users interactively specify their target shot arc, shoot the ball, and compare their shot arc to the target arc.}
    \label{fig:vis}
    \vspace{-2mm}
\end{figure*}

\section{System Design}
\label{sec:system}
We implemented two systems for basketball free-throw training (situated AR and co-located 2D display). Both systems share the same computer vision tracking component, but differ in their display mechanisms and visualizations.

\noindent
\textbf{Shot detection and tracking.}
Our system centers on the visualization of a player's actual shot arc compared to a target shot arc.
To define the target shot arc, users only need to specify the ball's launch angle and release point, since the arc endpoint is already given by the hoop.
To capture a player's shot arc, we implemented a shot tracking system that detects the ball's launch angle, release point, initial velocity, and endpoint at the rim (Fig.~\ref{fig:system}). 

Our shot tracking system is a real-time markerless 3D motion capture system consisting of four machine vision cameras (FLIR Imaging Systems) running at 80 fps. Two synchronized shooter cameras capture the initial ball motion upon ball release. 
Separately, two synchronized hoop cameras capture the ball's motion as it approaches the hoop and until the shot is over.  To maximize the contrast between the background and the basketball, we use white walls and floors in the background of each camera's view.
Simultaneously captured images from the synchronized cameras are sent frame-by-frame to a desktop machine. We threshold incoming images of the ball by brightness to detect pixels belonging to the ball. Next, we calculate the ball's position (in pixels) as the area centroid of the detected pixels within each camera's view. We then adjust this position to compensate for lens distortion and triangulate the ball's position in 3D world coordinates using calibrated intrinsic and extrinsic camera parameters. We repeat these steps for each consecutive pair of captured images to get the precise ball position for every time step.

The release point is the first available ball position upon release from the shooter's hand. 
We use tracking information of the 50 ms after ball release to calculate the ball's initial velocity and launch angle. 
The shot endpoint is extrapolated from the ball's tracked positions via the hoop cameras as it approaches the rim. 
We use approximately 250 ms of tracking data (20 measurements) before the ball touches or reaches the rim to extrapolate the ball's expected landing position, or shot endpoint. For our extrapolation, we assume that the ball reaches the 10 feet height of the rim without touching the rim, net, or backboard. 
Prior to our experiments, we performed an intrinsic calibration for each camera to compensate for distortion due to lens curvature and sensor misalignment. We also performed an extrinsic calibration between pairs of cameras to detect their relative position and orientation~\cite{brown_distortion, zhang_calibration_2000, tsai_3D_vision}. All image acquisition and tracking were implemented in Matlab~\cite{MATLAB:2016a}. 

Our intrinsic calibration error is approximately 0.007 pixels, which corresponds to approximately 14 micrometers ($\mu m$) of error in the undistorted ball position in a single camera (2D). The only ground truth available for each shot is that, according to physics, the ball's motion must be smooth from frame-to-frame. Thus, to measure the accuracy in tracking the ball's position in 3D, we quantify the high-frequency (frame-to-frame) noise in the ball's tracked position by fitting a 2nd order polynomial (constant acceleration) to the five position measurements over the 50 ms window after release for each shot. The standard deviation of the residuals from each fit represents the uncertainty in measuring the ball's position. From this, we calculate an average 3D tracking precision of $\sim$70 $\mu m$.
After tracking, we immediately transmit our shot tracking data to both the 2D and AR displays to create the arc visualizations within 100 ms.

Based on the tracking data, we can compute the shot arc visualization by using the projectile motion formula~\cite{walker_2017}, where \(
    x=v_0 t \cos(\theta)\) and
\(    
    y=v_0 t \sin(\theta) - \frac{1}{2}g t^2
\).
We obtain the full trajectory of $x$ and $y$ positions by considering the effect of gravitational acceleration ($g$), ball launch angle $\theta$, and the initial velocity $v_0$. 
Other factors, such as air resistance and ball spin, are omitted based on the assumption that they are negligible compared to gravity~\cite{chen2009physics}.

\noindent
\textbf{Co-located display with 2D visualization.}
Fig.~\ref{fig:vis}a shows the design and interaction methods of our 2D condition. We display a side-view of the shot arc on a typical 27" monitor (Fig.~\ref{fig:shooting_arc}a-2), which is directly connected to the desktop machine performing the tracking. We encode the spatial trajectory, ball launch angle, and release point into an arc visualization (\textbf{T1}). However, the co-located display does not afford the situatedness in the actual physical space. 
We designed our 2D system as the baseline condition in our comparative study, keeping the visualization and user input simple and easy to learn.
We set up the co-located display parallel to the shooter facing the hoop. This allows us to display the shot arc aligned to the real shooting direction, and minimizes the shooter's physical movement to access the visualization (\textbf{T2}).
We display both the target and actual shot arcs immediately after our tracking system has detected the shot outcome (\textbf{T3}). 
We encode the shot outcome as the color of the shot arc (i.e., green for a make and blue for a miss). Furthermore, we show the running score on the top-right of the screen.
Users can use the arrow keys on a keyboard to adjust the angle of the target shot arc at any time (\textbf{T4}).

\begin{figure*}[tb!]
    \centering
    \includegraphics[width=\linewidth]{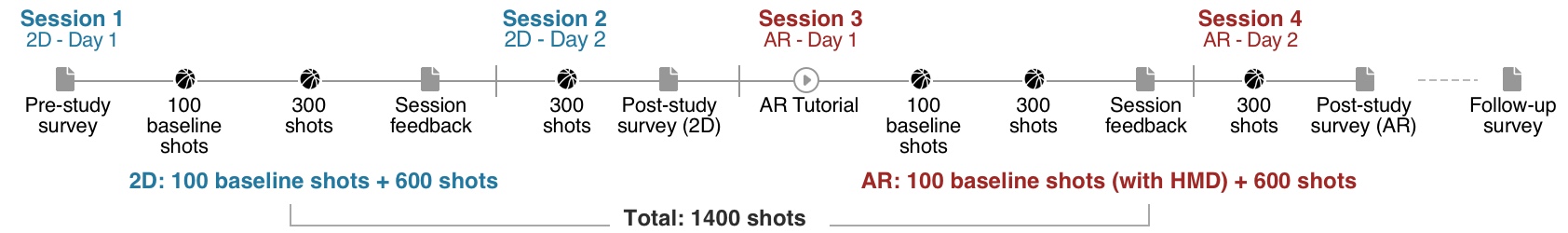}
    \caption{Experimental procedure. Each condition was measured in two consecutive sessions, with each session on a separate day. The starting condition was counterbalanced between 2D and AR. The study took four hours in total for each participant.}
    \Description[Experimental procedure.]{Experimental procedure. Each condition was measured in two consecutive sessions, with each session on a separate day. The starting condition was counterbalanced between 2D and AR. The study took four hours in total for each participant.}
    \label{fig:study_procedure}
\end{figure*}

\noindent
\textbf{AR HMD with situated 3D visualization.}
Fig.~\ref{fig:vis}b shows the design and interaction methods of our AR condition. We display situated 3D shot arcs projected directly into the player's physical space, leading from the shooter towards the hoop (\textbf{T2}, Fig.~\ref{fig:shooting_arc}b-2). 
We use a HoloLens (1st gen) and transmit our shot tracking data through WiFi. 
The hologram anchors the hoop's position and the position of the free-throw line and reconstructs the 3D shot arc based on the shooter's coordinates relative to the hoop center.
The main visual encoding is similar to the 2D condition and constitutes a shot arc, launch angle, and release point information. Additionally, we encode lateral spatial information based on the shot arc's three-dimensional embodied placement (\textbf{T1}). We display both the target and actual shot arcs and show the current score immediately after each shot is made (\textbf{T3}). 
We use the same color scheme as in the 2D condition, with careful selection of hue and opacity, to not distract the shooter. 
In the AR condition, we do not show the actual shot's launch angle in textual form because situated AR intrinsically supports shape understanding~\cite{St_John2001-cf}. Furthermore, we want to avoid potential information overload in AR~\cite{Pascoal2017-el}. 
Instead, users make a direct comparison of the shape of their target and actual shot arcs and adjust their angle and release point to refine their target arcs through gestural interaction (\textbf{T4}). 
Our hologram program runs at 60 fps to produce a stable visualization and minimize latency~\cite{hologram_stability}. Hologram stability in the HoloLens has been analyzed previously~\cite{vassallo2017hologram, guinet2019reliability} and Hologram drift under movement and sudden acceleration was determined to be 6.18mm on average.

Our 2D and AR systems are shown in Fig.~\ref{fig:shooting_arc} and in the supplemental video. Note that the AR shooter in the video looks at the shot arc instead of the hoop while shooting for demonstration purposes.


\section{Experiment Overview}
\label{sec:study}

Our study goal is to compare user training free-throw shooting with real-time co-located 2D and situated AR visual feedback.
We performed a controlled user study to collect qualitative user feedback and quantitative shot performance data with each condition.


\subsection{Experimental Conditions}

\noindent
\textbf{2D Condition.} 
In the co-located 2D visualization condition (Fig.~\ref{fig:vis}a), participants used the keyboard 
to adjust their target shot angle and fine-tune their horizontal shooting distance.
They then could use the visualization on the co-located screen to compare their target and actual shot arc with both the color-coded visual arc and the launch angle values on the screen.

\noindent
\textbf{AR Condition.}
In the situated AR visualization condition (Fig.~\ref{fig:vis}b), participants wore a HoloLens. They used gestures to adjust their target shot angle and ball release point. 
They then could look at the situated visualization in AR and compare the 3D arc shape of the target shot and actual shot after shooting.

\subsection{Experiment Set-Up}
\label{sec:set-up}

The study took place in an indoor laboratory setting with a designated area for free-throw shooting of approximate $6\times20$ feet (120 $ft^2$) in size, separated by a safety net from a working and rest area.
The hoop was installed at the standard 10 ft height, 15 ft from the free-throw line. Participants were instructed to stand behind the free-throw line to shoot, and an instructor stood under the hoop to rebound for participants. 

The monitor used in the 2D condition is a standard 27-inch monitor with a $1920\times1080$ resolution, and was placed on a 30-inch high desk with the keyboard, positioned to the participant's right, parallel to the shooting direction. 
Participants had to turn their heads about $75^\circ$ to look at the screen. 
The Hololens used in the AR condition has a field of view of $30^\circ\times17.5^\circ°$, a 60 Hz refresh rate, and weighs $579g$.
The instructor facilitated the initial spatial anchoring of the hoop location in the HoloLens.

\subsection{Participants}
We targeted experienced basketball players eager to improve their free-throw shooting.
We recruited 10 participants from Harvard school basketball clubs and university mailing lists. 

One participant was unable to complete the study due to the COVID-19 pandemic and has been excluded from the results. Among nine participants, three identified as female and six as male. They ranged in age from 18-35 years. Seven participants reported having prior experience with virtual reality without discomfort, but none had experience with AR HMDs. 

Five participants reported playing basketball for more than 10 years, three for 5-10 years, and one for 3-5 years. Seven participants reported playing at the intermediate or competitive level, and five of them actively played on school club teams or intramural teams. 
In terms of current training behaviors, seven participants played basketball more than once a week. Seven participants shot 100-300 shots during typical shooting practice, and the rest shot less than 100 shots. 
Seven out of nine participants reported knowledge about their shot percentages, but none knew their shot angles or target shot angle as they had no way of capturing them. Four participants reported that they kept their shooting records through oral counting or on paper.

\subsection{Design and Procedures}

We followed a within-subject experimental design to account for individual differences in free-throw skills. 
We tested each condition in two consecutive sessions, with 300 shots per session and 600 shots per condition. 
To compare performance improvement, we measure users' baseline performance through 100 shots prior to each condition. The number of free-throws in each session is based on our user interviews and a pre-pilot test with experienced shooters. The experienced shooters in our study have regularly practiced free-throws, with 100-300 shots in a typical practice, and tens of thousands of free-throws in their career. Therefore, we anticipate that without additional feedback from a coach or our tools their shot percentage will not improve. Subsequently, to minimize the physical burden on participants, we decided on measuring the baseline performance of participants in just 100 shots.

We counterbalanced the starting order of both conditions between participants.
The two sessions of the same condition were scheduled on separate days, one or two days apart, to avoid physical fatigue. 
Session 2 and 3 were scheduled a week apart to mitigate learning effects between the two conditions.

Here, we demonstrate our study procedure, starting with the 2D condition (Fig.~\ref{fig:study_procedure}).
In Session 1, we first introduced the purpose of the study and asked participants to sign a consent form. 
We instructed participants to \textit{use the tool to specify their desired angle and evaluate their shot arc within the visualization to improve free-throw shooting performance}. We emphasized that our goal was to evaluate the usefulness of our visualization design instead of their performance. 
%
Next, participants filled out a pre-study survey on their background and basketball experience. After doing sufficient warm-ups, they shot 100 baseline shots in their regular form.
We then introduced them to the 2D visualization tool. Participants were told to use the tool to adjust the target angle and compare the shot arc as frequently as they wanted during the training. The instructor was present to answer any questions. Next, participants continued to shoot 300 shots in the 2D condition. We encouraged them to take breaks between blocks (every 50 shots) and provided water. At the end of Session 1, we asked them for feedback about their training experience. In Session 2, they made another 300 shots with the 2D tool and filled out a post-study survey for the 2D condition. 

Session 3 was conducted about one week after Session 2. Due to the unfamiliarity of the HoloLens in our participants, they started with a gesture tutorial for the HoloLens. Next, participants shot 100 baseline shots in the AR condition, wearing a HoloLens without any visual feedback. 
We then introduced them to the AR visualization and let users get comfortable with it, which took roughly ten minutes. We told them to use the visualization as frequently as they wanted.
Next, participants shot 300 shots with the AR tool and gave session feedback in the end. In Session 4, they shot another 300 shots with the AR tool and filled out a post-study survey for the AR condition. They were asked to complete a follow-up survey two weeks after the study was complete. 
Overall, each session took roughly one hour to complete, and the whole study took about four hours per participant. We compensated participants with a \$40 Amazon gift card.

\subsection{Performance Metrics}
\label{sec:quantitative_measures}
We collected shot data from our shot tracking system and user inputs in the 2D and AR systems. In Sec.~\ref{sec:goal_task_analysis}, we identified \emph{improving shot arc and consistency with quantitative feedback to improve shot performance} as a main user goal. Therefore, we measure participants' shot angle, angle consistency, and shot percentage to capture training performance.
Below, we define the performance metrics we used to conduct the quantitative analysis.

\noindent \textbf{Initial baseline}: To evaluate how  user performance is impacted by the HMD, we compare the angle average of 100 regular baseline shots and 100 baseline shots while wearing the HMD (baseline+HMD).

\noindent \textbf{Shot angle consistency}: To compare how user shot angle performance has changed with training, we take the shot angle average and standard deviation of 100 regular baseline shots, and the last 100 shots in the 2D and AR conditions. 

\noindent \textbf{Shot accuracy}: To evaluate shot accuracy, we compare the shot percentage between 100 baseline shots, 100 baseline shots with HMD, and the last 100 shots in the 2D and AR conditions.

\subsection{Qualitative Feedback Collection}
\label{sec:qualitative_measures}
We collected subjective responses in session feedback, post-study surveys for both conditions, and a follow-up survey. 
To extract insights from the qualitative feedback, we performed qualitative coding 
among three coders. We categorized insights into a set of codes derived and agreed upon from an initial open coding 
by all three coders and applied the code set to the rest.
The complete survey questions with our coding results are provided in the supplemental material.
The following categories were analyzed.

\noindent \textbf{Visualization usefulness:} To characterize participants' training experiences, we analyze the open-ended user responses from session feedback and post-study surveys. These include their strategies to use the visualization, the insights derived from the visual feedback, and useful and missing aspects of the visualization to support their training.

\noindent \textbf{User goals:} To quantify the helpfulness of each visualization, we asked participants to rank their top goals from a list of goals (derived from our interviews). We also asked them to rate each tool accordingly in a follow-up survey. 

\noindent \textbf{Usability:} To evaluate the usability of both visualization tools, we analyzed the user ratings of 5-point Likert-scale questions from post-study surveys.

\section{Results and Discussion}
\label{sec:results}

This section presents the findings of our quantitative and qualitative analysis. We first focus on the observational aspects of our study and discuss the general user experience before analyzing training performance. Finally, we highlight unique aspects of AR and discuss tool usability.

\subsection{Real-time visual feedback is useful to refine subsequent shots}
\label{sec:visual_usefulness}
None of the participants had experienced free-throw shot training with real-time visual feedback before. 
To best capture individual and collective insights, we categorize user responses from the post-study surveys with representative quotations from participants and present the qualitative coding results.
We found that while the perceived usefulness of 2D and AR visualization varies by individual preference, real-time visual feedback helps users evaluate their performance and acquire self-generated insights.

\textbf{AR visualization provides useful guidance during shooting and helps generate insights on shooting form.}     
Using our AR condition, participants analyzed their previous shot arc, release point, and consistency, and focused on body form such as follow-through during shooting. \textit{“[I used] the 3D visualization to evaluate the shooting arc of the previous shot and using it as a reminder to follow through.”} (P1, AR).
\textit{”It was helpful to actually look and be able to visualize your shot angle and the ball going through the hoop.”} (P2, AR).
As a consequence, participants shifted their attention towards their shooting form and refined subsequent shots. \textit{“It forced me to feel the shot more than look at the rim, it forced me to arc higher.”} (P6, AR). \textit{“I like that it was able to analyze and help me fine-tune my release point.”} (P8, AR).
Participants also reported subjective improvement with the visual AR feedback. \textit{“[I] found it can help to enhance the muscle memory under a better shooting form and further improve the shot accuracy.”} (P5, AR).
\textit{“I think AR visualization was useful in guiding my shooting form, especially when I was more fatigued."} (P9, AR).
Collectively, 6 out of 9 participants commented on how the AR visualization led to improved form, either by allowing them to make informed decisions when shooting or by giving them specific insights on their form.
Six participants commented on the visual representation of the arc shape as one of the most useful aspects. Four participants also listed extra shot details such as direction, alignment, and exact release point as useful components. Three participants said they used the visualization to make explicit comparisons of aspects of their previous throws. 

\textbf{2D visualization allows direct comparisons with prior 
\newline
throws on angle performance.}
In 2D, users focused on analyzing shot angles \emph{after} their shot. 
\textit{“I think it is most helpful to analyze the shot angle after a missed shot to see what might have gone wrong.”} (P2, 2D). 
\textit{”I look at the arc after my shot. I think my optimal angle is 54.5. It helps me with consistency.”} (P3, 2D).
Participants derived insights on their performance to refine their shots. \textit{”It helped me learn a lot about my flaws - I seem to consistently make the same mistake of shooting at a smaller angle.”} (P3, 2D).
\textit{“I really liked seeing the instant feedback on my previous shot.  It actually helped me hone in on what was my ideal shot angle.”} (P6, AR).
Participants also noted differences between the two conditions. \textit{”The 2D visualization itself was not as helpful as the 3D one, but it was nice to pin down exactly what my target angle was.”} (P8, 2D). \textit{“I felt like 2D gave better immediate feedback on performance since I could easily compare my arc vs. my desired arc.”} (P6, 2D). Overall, participants found the numeric display of the angle number useful (8 out of 9). Four participants reported using the angle information to make after-the-fact comparisons with prior throws. Three participants also commented on the ease of use of the visualization.

\subsection{Each visualization modality naturally supports different
user goals}
\label{sec:user_goal}
\begin{figure}[t]
    \centering
    \includegraphics[width=\linewidth]{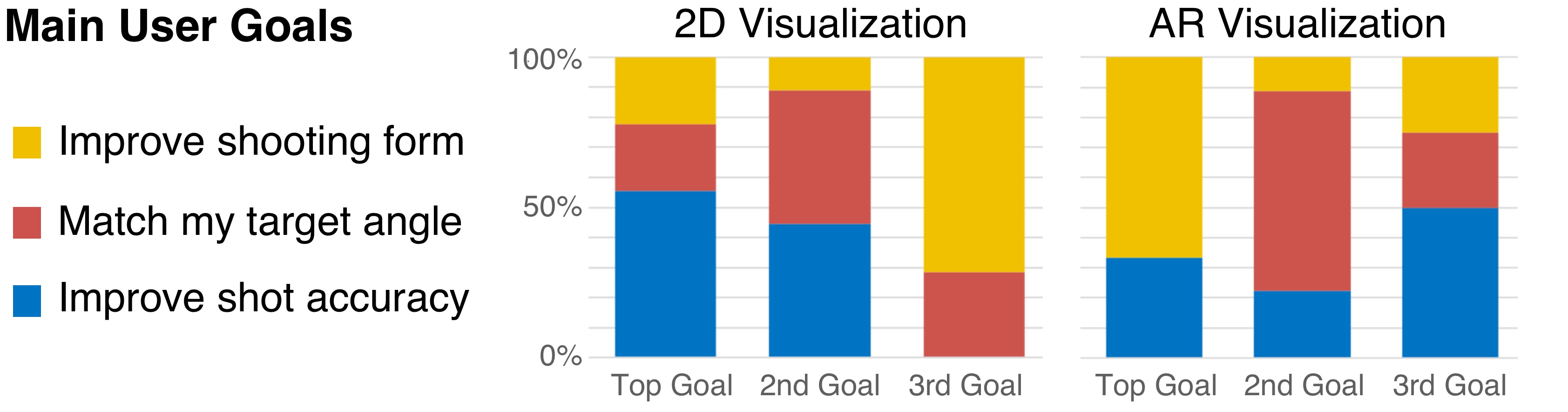}
    \caption{Main user goals of the 2D and AR visualizations.}
    \Description[Main user goals of the 2D and AR visualizations.]{Main user goals of the 2D and AR visualizations.}
    \label{fig:survey_response_goals}
    \vspace{-2mm}
\end{figure}

In the follow-up survey, we asked participants to rank their top goals when using the AR and 2D systems
among \textit{``Improve shooting form'', ``Match target angle'' and ``Improve shot accuracy''}.
We found that the majority of AR users (6) indicated their main goal was to improve shooting form, while most 2D users (5) reported their main goal was to improve shot accuracy (Fig.~\ref{fig:survey_response_goals}). The remaining AR users (3) reported their main goal was to improve shot accuracy, and the remaining 2D users spread evenly between improving form (2) and matching their target angle (2). 

We asked participants to rate how helpful each tool was in reaching their goals from 1 (not helpful at all) to 5 (very helpful).
Within participants who reported that improving form was their top goal, 4 of 6 participants rated the AR system as 4 or 5 for helpfulness in achieving their goals. These participants reported that it aided their consistency and muscle memory. The remaining 2 of 6 participants gave ratings of 2, citing the headset's discomfort as a major detractor. For the 2D system, both participants aiming to improve their form gave ratings of 3. 

Within participants who aimed to improve shot accuracy as their top goal, 3 out of 3 gave the AR system a rating of 3 or 4 on the helpfulness scale, citing improvements in their release point or arc height. For the 2D system, 4 out of 5 participants gave the system a rating of 4 or 5 on the helpfulness scale. Participants particularly appreciated the ability to make a numeric comparison between their target and actual shooting angles. At the same time, participants reported that the 2D system did not help them improve their overall form and did not allow them to fine-tune other aspects of the feedback, such as the release point. 

Taken together, we see that our AR and 2D systems differ in which goals they promote within users, and in how helpful they are in supporting those goals. Particularly, \textbf{the AR system appears to promote a focus on improving shooting form}, for which it was ranked more helpful than the 2D system. \textbf{The 2D system, however, appears to promote a focus on improving shot accuracy}, for which it was ranked more helpful than the AR system.

\begin{figure*}[bt!]
    \centering
    \includegraphics[width=\linewidth]{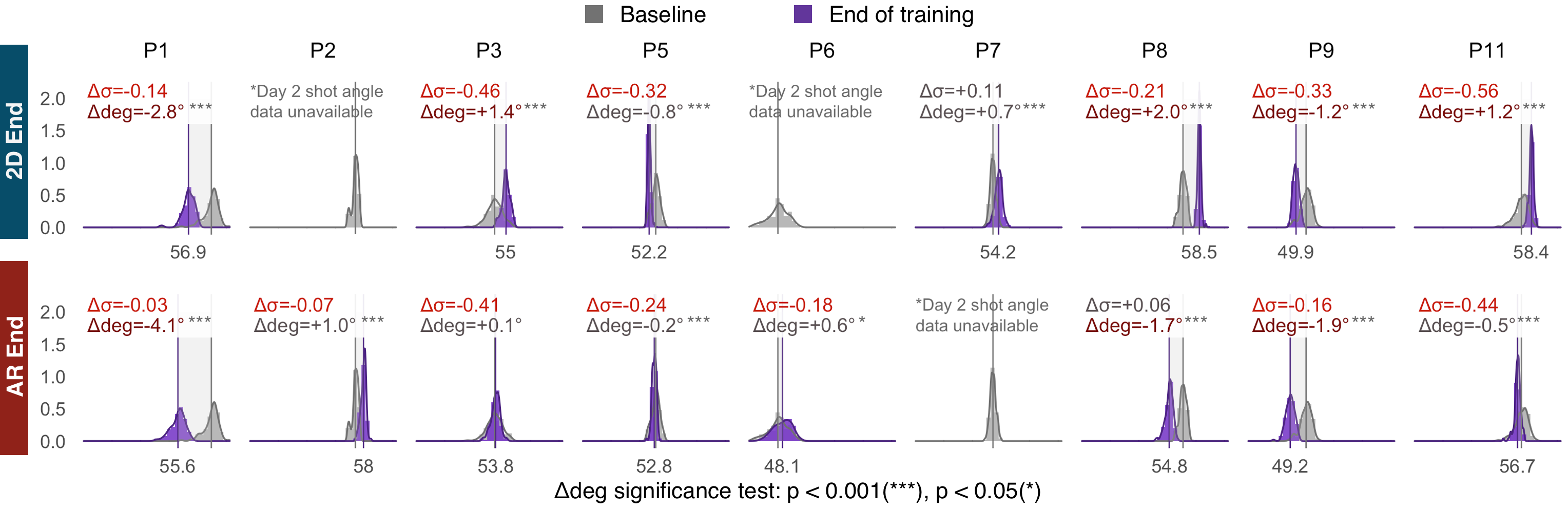}
    \caption{Players' shot angle distribution plots, comparing their baseline performance to their end of training performance in the 2D (top row) and AR (bottom row) condition, respectively. Each panel shows the difference in standard deviation ($\Delta \sigma$) and angle ($\Delta$deg). 
    Comparing standard deviation ($\Delta \sigma$), most participants (7/8 in AR and 6/7 in 2D) improved on angle consistency (p<0.05). Comparing shot angle ($\Delta$deg), most 2D users (5/7) changed their angle after training ($\Delta$deg$>1^\circ$) while most AR users (5/8) shot at the same angle ($\Delta$deg$\leq1^\circ$). }
    \Description[Players' shot angle distribution plots, comparing their baseline performance to their end of training performance in the 2D (top row) and AR (bottom row) condition, respectively.]{Players' shot angle distribution plots, comparing their baseline performance to their end of training performance in the 2D (top row) and AR (bottom row) condition, respectively. Each panel shows the difference in standard deviation ($\Delta \sigma$) and angle ($\Delta$deg). 
    Comparing standard deviation ($\Delta \sigma$), most participants (7/8 in AR and 6/7 in 2D) improved on angle consistency (p<0.05). Comparing shot angle ($\Delta$deg), most 2D users (5/7) changed their angle after training ($\Delta$deg$>1^\circ$) while most AR users (5/8) shot at the same angle ($\Delta$deg$\leq1^\circ$). }
    \label{fig:base_end}
\end{figure*}

\begin{figure*}[bt!]
    \centering
    \includegraphics[width=\linewidth]{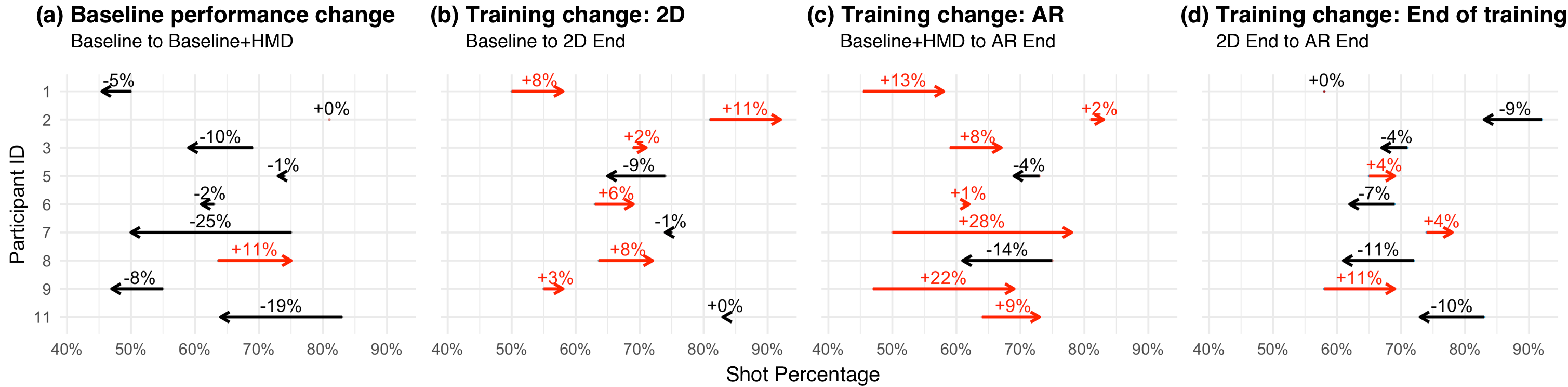}
    \caption{Players' shot percentage comparison before and after training. For initial performance (a), 7/9 participants had lower performance wearing an HMD. Between end of training and baseline performance, most reached a higher shot percentage, 6/9 in 2D (b), and 7/9 in AR (c). Between 2D and AR end performance (d), more perform better in 2D (5) than in AR (3). }
    \Description[Players' shot percentage comparison before and after training.]{Players' shot percentage comparison before and after training. For initial performance (a), 7/9 participants had lower performance wearing an HMD. Between end of training and baseline performance, most reached a higher shot percentage, 6/9 in 2D (b), and 7/9 in AR (c). Between 2D and AR end performance (d), more perform better in 2D (5) than in AR (3).}
    \label{fig:shot_percentage}
\end{figure*}

\subsection{Performance in shot angle consistency improved throughout the study}
\label{sec:quantitative}
We analyzed shot performance of our participants before and after training, including shot angle and consistency. Our hypothesis was that training with real-time visual feedback would improve participants' shot arc consistency, which can further lead to shot percentage improvement. 
 We used the Wilcoxon signed-rank test to check for significance in shot angle comparison, angle consistency, and shot percentage improvement. 
In Fig.~\ref{fig:base_end}, we show the changes in shot angle consistency ($\Delta\sigma$) and angle ($\Delta$deg) before and after training. 
We show the shot angle distribution of 100 regular baseline shots compared to the end of training for the 2D (top row) and AR (bottom row) condition.
Due to tracking errors, a portion of angle data is unavailable for P2, P6, and P7.
We interpret the r effect size using Cohen's classification, which is 0.1, 0.3, and 0.5 and above for small, moderate, and large effects, respectively~\cite{cohen2013statistical}. 
Among the available records, 6 out of 7 participants in 2D increased their angle consistency ($\Delta \sigma$=-0.27, r effect size= 0.83, p=0.03). 7 out of 8 participants in AR increased their angle consistency ($\Delta \sigma$=-0.19, r effect size=0.83, p=0.02). 
\textbf{The results show significant improvement of angle consistency in both AR and 2D after training with visual feedback}, which complies with our hypothesis.

Fig.~\ref{fig:shot_percentage} shows a comparison of shot percentages of all participants between their baseline shots and end of training in 2D and AR. To understand the impact of shooting with a headset, in Fig.~\ref{fig:shot_percentage}a, we compare the baseline performance change between regular baseline (i.e., shooting without any visual feedback) and baseline+HMD (i.e., shooting without any visual feedback but while wearing a headset). 7 out of 9 participants had a drop in shot percentage due to wearing a headset. 
Fig.~\ref{fig:shot_percentage}b and c show the performance change from start of training to end of training, for the 2D and AR conditions, respectively.
6 out of 9 in 2D and 7 out of 9 in AR increased their shot percentage. 
Fig.~\ref{fig:shot_percentage}d shows performance differences at the end of training between 2D and AR.
Despite initially being distracted by the headset, 3 out of 9 participants had a higher shot percentage in AR, while 5 performed better in 2D. However, \textbf{there is no significant improvement among all participants' shot percentage in both conditions}. This is likely due to the short-term exploratory design of our experiment, as training sports skills requires repetitive practice over a longer time. Another observation is the negative impact on initial shot performance from wearing an HMD. We discuss the impact of wearing an HMD in more detail in Sec.~\ref{sec:ar_adjustment}.

\begin{figure*}[bt!]
    \centering
    \includegraphics[width=\linewidth]{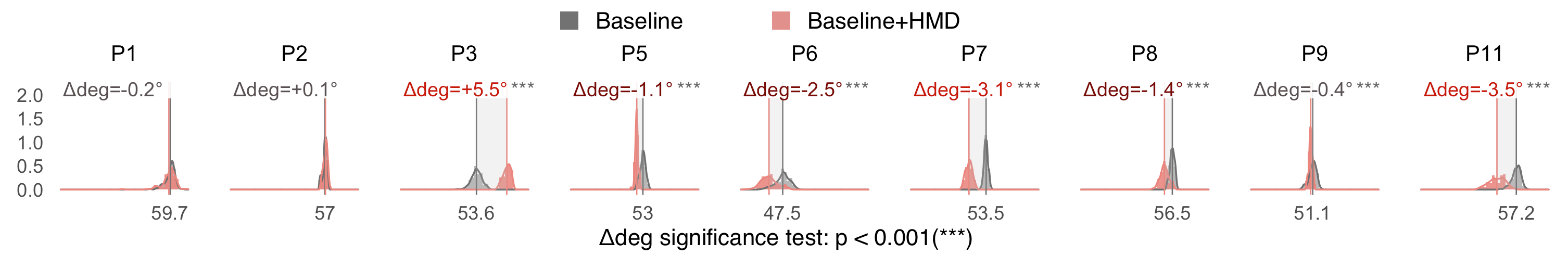}
    \caption{Shot angle comparison between 100 baseline shots (Baseline) and while wearing an HMD without visual feedback (Baseline+HMD). 6 out of 9 participants had an angle change ($\Delta$deg$>1^\circ$). P3, P7, and P11 had a larger than $3^\circ$ angle change, with the most initial shot percentage decrease in Fig.~\ref{fig:shot_percentage}. }
    \Description[Shot angle comparison between 100 baseline shots (Baseline) and while wearing an HMD without visual feedback (Baseline+HMD).]{Shot angle comparison between 100 baseline shots (Baseline) and while wearing an HMD without visual feedback (Baseline+HMD). 6 out of 9 participants had an angle change ($\Delta$deg$>1^\circ$). P3, P7, and P11 had a larger than $3^\circ$ angle change, with the most initial shot percentage decrease in Fig.~\ref{fig:shot_percentage}.}
    \label{fig:initial_angle}
\end{figure*}

\subsection{Immersive visualization requires a higher level of adjustment}
\label{sec:ar_adjustment}
The shot data and user feedback indicate that wearing an HMD requires varying degrees of adjustment. The immersive visualization also requires some adaptation. However, the impact on shot performance and user perception was mitigated over time.    

Here, we investigate the impact of wearing an HMD on the initial shot angle and shot performance. 
Fig.~\ref{fig:initial_angle} shows that 6 out of 9 participants had a noticeable change in their shot angle ($>1^{\circ}$, p<0.001) in baseline+HMD, 3 of which (P3, P7, P11) had a massive change ($>3^{\circ}$). Notably, they also had the sharpest decline in shot percentage compared to their baseline performance  ($-10\%, -25\%, -19\%$, respectively), shown in Fig.~\ref{fig:shot_percentage}.
This shows that the impact of the AR headset varies by individuals. Nevertheless, over the 2-day training period, \textbf{most AR participants improved on shot angle consistency (7 out of 8) and outperformed their initial shot percentage (7 out of 9)}.

From our qualitative coding of the post-study surveys, 6 out of 9 participants cited the discomfort and intrusiveness of the headset as a negative aspect. They also said it took some time to adapt to the headset. \textit{“In the first two rounds, the headset was distracting, especially when it was slipping off. It took a bit of time to get used to it.”} (P1, AR). \textit{“At the beginning, I didn't feel too comfortable with this. [The] device is on my shooting path, so I have to change my posture to accommodate the device.”} (P11, AR). Users had to get used to the situated AR visualization due to its unfamiliarity but appreciated the benefits of an AR visualization. 
\textit{“The visualization took a bit to get used to but was interesting to analyze previous shots or any that did not feel right on release.”} (P8, AR).
\textit{“Nice not to
have to wear the headset [in 2D], but the visualization in AR is easier to use.”} ( P2, 2D).

As the physical presence and associated discomfort of the HoloLens was the biggest detractor from the AR condition, a more lightweight headset should be used for future sports AR design. \textbf{After initial adaptation, most participants showed improvement with AR visual feedback during the study}. Conducting a longer-term study is likely to lead to more measurable improvement in shooting performance.

\subsection{Immersive visualization resulted in a more holistic sense of participant performance}
\label{sec:holistic}
As shown in Fig.~\ref{fig:survey_response_goals}, user goals differ in the AR and 2D conditions. The AR condition promotes a focus on body form compared to the emphasis on shot accuracy in the 2D condition. Our quantitative angle data and qualitative feedback both support this observation.

In Fig.~\ref{fig:base_end}, we investigate how the shot angle of participants changed ($\Delta$deg) from their baseline shots to the end of their training. We found that most participants in AR (5 out of 8) stayed at a constant angle ($|\Delta$deg$|\leq1^\circ$), while in 2D, most participants (5 out of 7) ended up adjusting their shot angle ($|\Delta$deg$|>1^\circ$). \textbf{This reflects the focus on the overall form instead of a specific angle in the AR condition.} 

From the user feedback on the visualization's usefulness, AR users commented that situated visualization feels more realistic and provides useful additional information, such as the ball's release point. \textit{“The visualization felt more realistic with the 3D tool.”} (P1, AR).  Four participants cited extra shot details such as shot alignment and exact release point as useful.
\textit{”I can see the release point in the AR. It also visualized my shooting path, which kept reminding me [of] my form.”} (P11, AR). On the other hand, some participants feel the AR visualization cause overthinking. \textit{“Sometimes I think it might cause me to overthink my shot, however. Still, it definitely makes me more aware of the way in which I am releasing the ball and at what angle I am releasing it.”} (P2, AR). 
Participants found the 2D visualization to be limited for evaluating their shots more holistically and were able to extract more shot details, such as direction, alignment, and release point from the AR visualization.
\textit{“I think it is useful to evaluate the shot angle, but it seems not very useful for the body form.”} (P9, 2D).
\textit{“The visualization itself was not as helpful as the 3D one. This doesn't allow for corrections in the release point like the 3D visualization.”} (P8, 2D).

Taken together with the results from Sec.~\ref{sec:visual_usefulness}, \textbf{situated AR visualization encourages training with a focus on body form and provides more holistic feedback on a player's performance. It gives useful guidance during shooting and supports specific insight generation.} Users also indicated that the highly situated nature of an AR display makes it increasingly important to
only display relevant information to avoid sensory overload. To accomplish this, interface designers should tailor the visual feedback to the user's specific goals.

\subsection{Lighter headsets will likely increase the real-world applicability of AR training}

\begin{figure}[t!]
    \centering
    \includegraphics[width=\columnwidth]{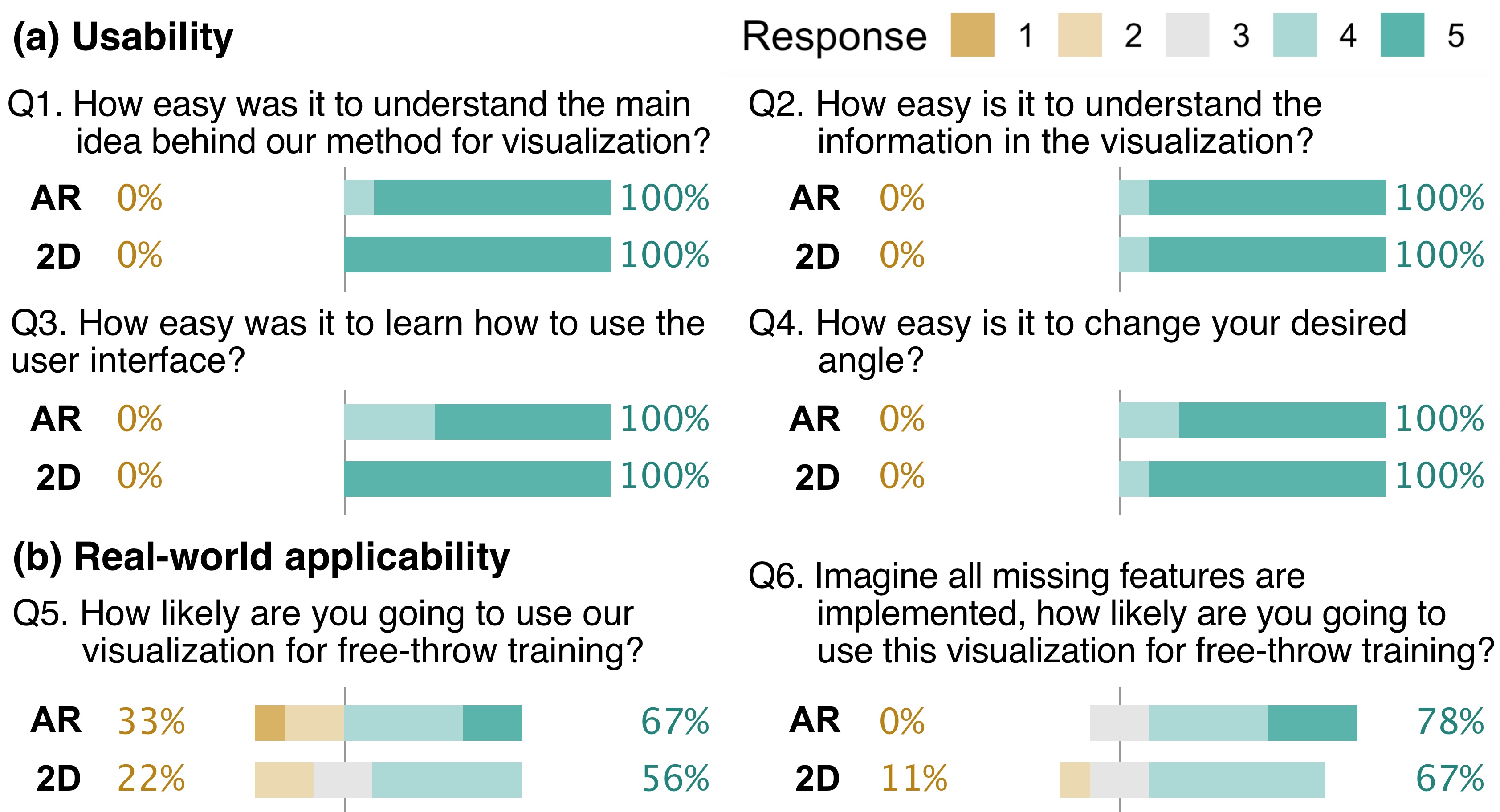}
    \caption{Tool usability ratings. The percentage number on the left represents negative responses (1 and 2) in beige, and the percentage on the right represents positive responses (4 and 5) in teal. Neutral responses (3) are not included.}
    \Description[Tool usability ratings.]{Tool usability ratings. The percentage number on the left represents negative responses (1 and 2) in beige, and the percentage on the right represents positive responses (4 and 5) in teal. Neutral responses (3) are not included.}
    \label{fig:survey_response}
    \vspace{-2mm}
\end{figure}

We want to understand participants' acceptance and preference on our shot arc visualization approaches, as they are considered novel for shot training. In Fig~\ref{fig:survey_response}, we asked for user ratings on both tool usability and real-world applicability. 
All nine participants rate questions Q1 to Q4 positively, reporting the ease of understanding the visualization and learning to use the tool for both the AR and 2D conditions. 
The high average ratings for all questions combined in both AR ($\mu$=4.8) and 2D ($\mu$=4.9) reflect the high usability of the visualization and tools. 
In Fig~\ref{fig:survey_response}b, we ask how likely participants are to user our tools in real training (Q5).
Of all 9 participants, 67\% rate AR positively ($\mu$=3.4) and 56\% rate 2D positively ($\mu$=3.3), while 33\% in AR and 22\% in 2D rate it negatively. 
In Q6, we report participants' willingness to use the tools after adding missing features (less intrusive HMD in AR and extra shot details in 2D as described in Sec.~\ref{sec:holistic}).
78\% in AR ($\mu$=4.1) and 67\% in 2D ($\mu$=3.6) report positively on this. 
It is worth noting that 22\% and 33\% of participants in Q5 and Q6, respectively, rate AR with 5s, while nobody rated the 2D condition with 5s. 
From Q5 to Q6, the responses in AR also noticeably shift from 33\% negative responses to no negative responses. 
\textbf{The results show that the AR condition received more contrasting ratings than the 2D condition, and became more favorable than 2D if the missing features, such as a lighter headset weight, were to be improved.}

\section{Implications for Sports XR}
\label{sec:discussion}

We provide a set of design considerations for AR sports skill training tools and study design. These considerations include device recommendations, study goal settings, and player-related guidelines. 

The largest hurdle in the AR condition was the headset itself. The headset was heavy and bulky, required frequent re-positioning for some participants, and provided a limited FoV. The limitations imposed by the headset can be greatly mitigated by using the increasingly lightweight AR devices that are entering the market, such as Nreal AR glasses~\cite{nreal} that weighs only $88g$.
When designing future studies and analyzing participants' performance results, we need to consider the impact of wearing an AR headset on participants' performance. \textbf{We believe a lightweight device can mitigate the negative impact of wearing a headset on participants' performance}.

One of our major takeaways is the importance of giving athletes clear instructions on a system's training goal. 
Fig.~\ref{fig:survey_response_goals} shows that participants in our study had a wide range of goals they targeted.
Some goals (e.g., shot angle vs. shot percentage) might be competing goals, where, in the short run, changing one might negatively impact the other. 
This is common in sports skill training, where athletes often see an initial decrease in overall performance when changing a specific aspect of their technique.
While the open-ended nature of the current study allowed us to better understand how participants naturally interact with the different interfaces, \textbf{a future study that controls for participant goals will likely be able to draw more conclusive insights in regards to actual improvements that are facilitated by different interfaces.} 

Finally, the targeted user group's skill level plays a large role in the design of the training system.
Player skill levels and prior experiences impact the types of feedback needed. 
The basketball coaches in our interviews confirmed that they often tailor their feedback based on a player's skill level.
A practical consequence of this is that a single training system design will likely not lead to improvement for all players, and experienced players will likely not exhibit an improvement over the relatively short time span afforded in a typical user study. 
This is reflected in the diverse user response on visualization usefulness in Sec.~\ref{sec:visual_usefulness} and performance metrics observed in Sec.~\ref{sec:quantitative}. 
\textbf{Future studies would benefit from tailoring the types and amount of visual feedback to the athlete's skill level}. 
Furthermore, within the scope of our study, we have focused on the benefits of real-time visual feedback but have not investigated whether the gained skills are transferable. Evaluating the post training performance without any feedback from the tool is necessary to support long term skill learning~\cite{anderson2013learning}.
\textbf{Future studies should also evaluate motor-skill improvements over a larger time span and adopt adaptive guidance in the system design.}


\section{Conclusions and Future Work}

This paper aims to provide insights for applying co-located 2D and situated AR visualizations to realistic sports skill training of basketball free-throw shooting with immediate visual feedback. 
Results suggest that, first, real-time visual feedback supports users to refine subsequent shots, and each visualization modality supports distinct user goals. Second, user shot angle consistency improved with both 2D and AR visualizations. Third, despite the initial impact of wearing an AR headset, immersive visualization facilitates a more holistic sense of performance and body form in athletes.  
Overall, players favored AR in real-world training if the headset discomfort were to be improved. 

As the first comprehensive study on applying situated AR visualization to basketball free-throw training, our insights set a starting point for future research in immersive sports skill training.
For future work, identifying specific training goals based on sports type and skill levels, and designing tailored visual feedback with consideration of the guidance hypothesis for motor learning~\cite{schmidt1991frequent}, are considerable HCI research challenges.
We also envision integrating body tracking measures to the visual feedback. 
Finally, Sports AR can provide personalized quantitative visual feedback at low cost, and opens up exciting new opportunities for remote collaboration between coaches and athletes.


\begin{acks}
We wish to thank coaches and players on Harvard Women’s and Men’s Basketball teams for their time and expertise. 
This research is supported in part by the National Science Foundation (NSF) under NSF Award Number IIS-1901030, and the Harvard Physical Sciences and Engineering Accelerator Award.
\end{acks}

\bibliographystyle{ACM-Reference-Format}
\bibliography{main}


\end{document}